\documentclass[preprint,12pt]{elsarticle}

\usepackage{amssymb}
\usepackage{amsmath}
\usepackage{bm}
\usepackage{comment}

\usepackage{xcolor}

\usepackage[normalem]{ulem}  % \sout{old text} for strikeout

\renewcommand\sout{\bgroup \color{red} \ULdepth=-.5ex \ULset}

\begin{document}
\begin{frontmatter}



\title{Theoretical study on $\Lambda\alpha$ and $\Xi\alpha$ correlation functions}


\author[1]{Asanosuke Jinno} %% Author name
\ead{jinno.asanosuke.36w@st.kyoto-u.ac.jp}

\affiliation[1]{organization={Department of Physics, Faculty of Science},%Department and Organization
            %addressline={}, 
            city={Kyoto},
            postcode={606-8502}, 
            country={Japan}}

\author[2,3,4]{Yuki Kamiya} %% Author name
\ead{yuki.kamiya.d3@tohoku.ac.jp}

\affiliation[2]{organization={Helmholtz Institut für Strahlen- und Kernphysik and Bethe Center for Theoretical Physics, Universität Bonn},%Department and Organization
				%addressline={}, 
				city={Bonn},
				postcode={D-53115}, 
				country={Germany}}

\affiliation[3]{organization={RIKEN Interdisciplinary Theoretical and Mathematical Science Program (iTHEMS)},%Department and Organization
				%addressline={}, 
				city={Wako},
				postcode={351-0198}, 
				country={Japan}}

\affiliation[4]{organization={Department of Physics,
Tohoku University},%Department and Organization
				%addressline={}, 
				city={Sendai},
				postcode={980-8578}, 
				country={Japan}}

\author[5,3]{Tetsuo Hyodo} %% Author name
\ead{hyodo@tmu.ac.jp}

\affiliation[5]{organization={Department of Physics, Tokyo Metropolitan University},%Department and Organization
				%addressline={}, 
				city={Hachioji},
				postcode={192-0397}, 
				country={Japan}}

\author[6]{Akira Ohnishi\corref{cor1}}
\cortext[cor1]{Deceased.}
\affiliation[6]{organization={Yukawa Institute for Theoretical Physics, Kyoto University},%Department and Organization
				%addressline={}, 
				city={Kyoto},
				postcode={606-8502}, 
				country={Japan}}

%% Abstract
\begin{abstract}

We examine the $\Lambda$-${}^4\mathrm{He}$ ($\alpha$) and $\Xi \alpha$ momentum correlation in high-energy collisions to further elucidate the properties of the hyperon-nucleon interactions.
For the $\Lambda\alpha$ system, we compare  $\Lambda\alpha$ potential models with different strengths at short range.
We find that the difference among the models is visible in the momentum correlation from a small-size source.
This indicates that the $\Lambda\alpha$ momentum correlation can constrain the property of the $\Lambda N$ interaction at short range,
which plays an essential role in dense nuclear matter.
For the $\Xi\alpha$ system, we employ the folding $\Xi \alpha$ potential based on the lattice QCD $\Xi N$ interactions.
The $\Xi \alpha$ potential supports a Coulomb assisted bound state of ${}^5_{\Xi}\mathrm{H}$ in the $\Xi^-\alpha$ channel, while the $\Xi^0\alpha$ channel is unbound.
To examine the sensitivity of the correlation function to the nature of the $\Xi \alpha$ potential, we vary the potential strength simulating stronger and weaker interactions.
The result of the correlation function clearly reflects the bound state nature in the $\Xi^-\alpha$ correlation.

\end{abstract}

%% Keywords

\end{frontmatter}



\section{Introduction}
\label{sec:Intro}
%% Labels are used to cross-reference an item using \ref command.

Revealing the nature of the hyperon ($Y$)-nucleon ($N$) interactions is essential for enhancing our understanding of various physics including neutron stars, heavy-ion collision reactions, hypernuclei, and dibaryons.
The properties of the hyperon-nucleon interactions have been examined based on the hypernuclear spectroscopy, the hyperon-nucleon scattering data, and neutron star observations.
Nevertheless, due to the insufficient constraint on the $YN$ interactions, there are still unresolved problems such as the hyperon puzzle of neutron stars~\cite{Demorest:2010bx,Gal2016,Burgio:2021vgk} and the existence of the $H$ dibaryon~\cite{Jaffe:1976yi,HALQCD:2019wsz,Gal:2024nbr}.

Recently, the momentum correlation created in high-energy collisions has been intensively investigated to obtain further information on the baryon-baryon interactions (see Refs.~\cite{Jinno:2024tjh,Kamiya:2024diw} and references therein).
These studies are called femtoscopy.
Over the past few years, the momentum correlation function of the mass number $A=3$ systems garners interest as it reveals additional properties of the two-body interactions and to access the many-body interaction~\cite{ALICE:2023bny}.

The $Y$-$^{4}{\rm He}$ ($\alpha$) momentum correlaion can be a good probe to obtain further information on the $YN$ and possibly $YNN$ interactions for several reasons.
The central density inside $\alpha$ can reach $2\rho_0$~\cite{Hofstadter:1957wk}, and then the property of the hyperon in dense nuclear matter may be obtained through constraining the $Y\alpha$ potential.
Since both the total spin and isospin of $\alpha$ are zero, the $Y\alpha$ momentum correlation can be calculated by using only a single-channel $Y\alpha$ potential, in contrast to the $YN$ correlations with multiple spin components.
This provides an advantage when comparing to experimental measurements, where only the spin-summed correlation can be measured.

In this proceeding, we investigate the momentum correlations of the $\Lambda\alpha$~\cite{Jinno:2024tjh} and $\Xi\alpha$~\cite{Kamiya:2024diw} systems.
Each system has its own motivation: The $\Lambda\alpha$ system is examined to constrain the short range part of the $\Lambda N$ interactions and the property of $\Lambda$ in dense nuclear matter.
The $\Xi\alpha$ system is explored to pin down the unknown bound state properties of the $\Xi \alpha$ systems.
In Sec.~\ref{sec:CF}, we describe the method to calculate the momentum correlation.
The background, models, and results of the $\Lambda \alpha$ and $\Xi \alpha$ systems are given in  Secs.~\ref{sec:LamAlpha} and \ref{sec:XiAlpha}, respectively.
The conclusions and outlook are given in Sec.~\ref{sec:summary}.

\section{Correlation function with $\alpha$}
\label{sec:CF}

In this study, we consider the $Y\alpha$ momentum correlation created in high-energy collisions.
We assume the $Y\alpha$ system as a two-body system because $\alpha$ is tightly bounded object.
The many-body nature of $\alpha$ and the $\alpha$ break-up processes 
are not explicitly considered.
To calculate the correlation function $C$ as a function of the relative momentum $\bm{q}$, we employ the Koonin-Pratt formula~\cite{Koonin:1977fh,Pratt:1986cc,ExHIC:2017smd}
\begin{align}
	C(q) = \int d\bm{r} S(r) \left|\Psi^{(-)}(\bm{r},\bm{q})\right|^2,
	\label{eq:KPformula}
\end{align}
where we assume the source function $S$ as a spherical and static Gaussian source function $S(r)=\exp{(-r^2/4R^2)}/{(4\pi R^2)^{3/2}}$ with the source size $R$.
For the relative wave function of the $Y\alpha$ system $\Psi^{(-)}$ with the outgoing boundary condition, we include the strong $Y\alpha$ interaction effect only in the $s$-wave state since the sizable correlation emerges only in the low-momentum region.
When there is the Coulomb interaction, one must use the Coulomb wave functions as the asymptotic form in all the partial waves, and introduce the strong interaction effect in the $s$ wave~\cite{Kamiya:2019uiw}.

We will treat the source size of $1~\text{fm}$, which is smaller than the charge radius of $\alpha \approx 1.7~\text{fm}$~\cite{Krauth:2021foz}.
This may seem too small, since the hyperon would be apparently produced inside $\alpha$.
Actually, however, the probability distribution of the relative distance $r$ is given by $4\pi r^2 S(r)$~\cite{Mihaylov:2018rva}. Then, the mean distance between the emitted pair is given by $\langle r\rangle=4R/\sqrt{\pi}\approx 2.26~\text{fm}$ for $R=1~{\rm fm}$.

\section{$\Lambda \alpha$ momentum correlation}
\label{sec:LamAlpha}

Properties of the $\Lambda \alpha$ potential have been investigated in several studies.
The overall attraction is constrained by the ${}^5_\Lambda {\rm He}$ ($\Lambda + \alpha$) binding energy data of $3.12~{\rm MeV}$~\cite{Juric:1973zq}.
The approximate interaction range can be estimated as the sum of the radius of $\alpha$ and the range of the two-pion exchange.
However, little is known about the short-range behavior of the $\Lambda \alpha$ potential, which is nevertheless important for understanding the properties of $\Lambda$ in dense nuclear matter.
The study of the light $\Lambda$ hypernuclei~\cite{Motoba:1985uj} shows that the calculated binding energy is not sensitive to the variation in the short-range behavior.
The weak-decay width of the light $\Lambda$ hypernuclei is better reproduced with the repulsive core than with the attractive one~\cite{Kumagai-Fuse:1994ulj}.
Then, it is beneficial to consider whether the $\Lambda \alpha$ momentum correlation could provide another perspective on the short-range part of the $\Lambda\alpha$ potential.

\begin{figure}[t]
\centering
\includegraphics[scale=0.4]{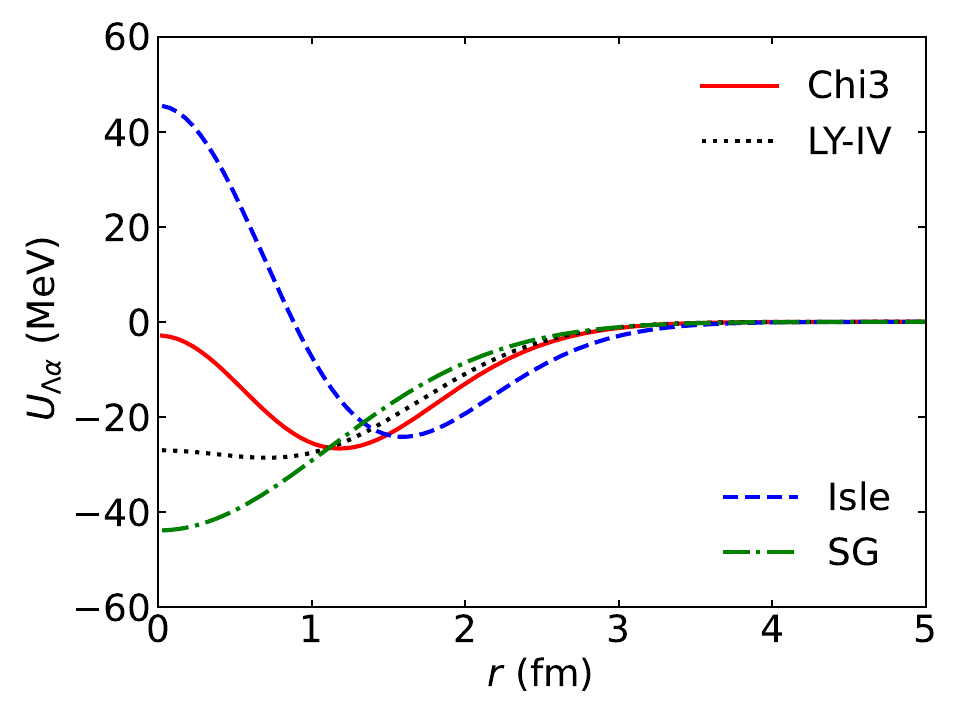}
\includegraphics[scale=0.4]{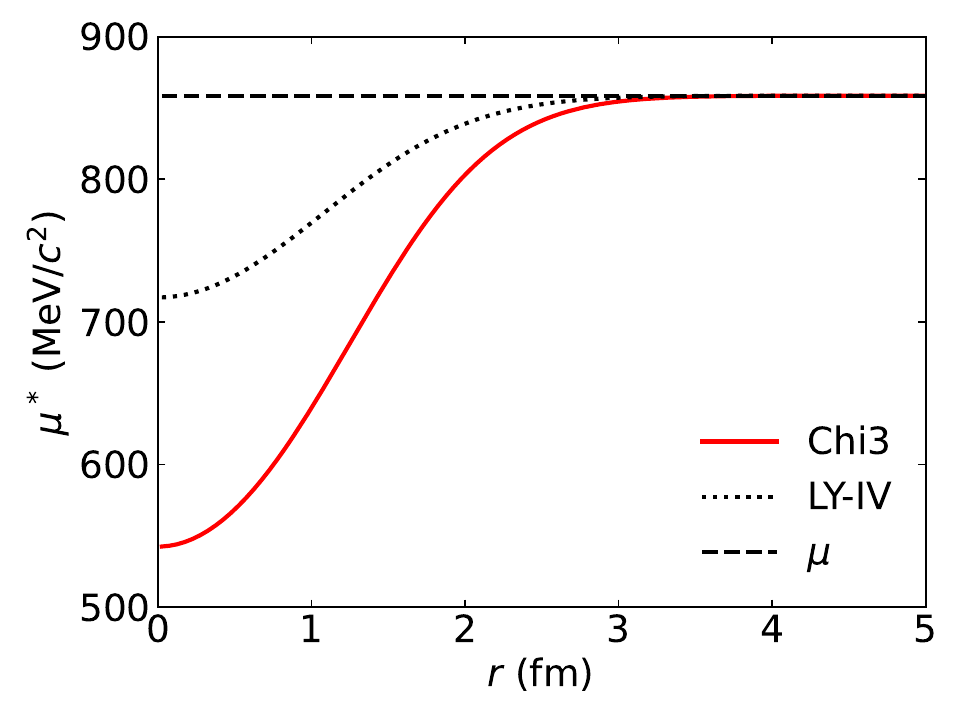}
\caption{$\Lambda\alpha$ potentials (left panel) and reduced effective masses (right panel) as functions of the distance between $\Lambda$ and $\alpha$.
Chi3 (solid line) and LY-IV (dotted line) are based on the Skyrme-type $\Lambda$ potentials~\cite{Jinno:2023xjr} with the $\alpha$ density distribution.
Isle (dashed line in the left panel) and SG (dash-dotted line) are the phenomenological models~\cite{Kumagai-Fuse:1994ulj}.
The dashed line in the right panel represents the reduced mass in vacuum of the $\Lambda \alpha$ system.
Figures are adopted from Ref.~\cite{Jinno:2024tjh}.}
\label{fig:LamAlphaPot}
\end{figure}

We compare four $\Lambda\alpha$ potential models with different strength at short range.
Isle and SG potentials~\cite{Kumagai-Fuse:1994ulj} are the phenomenological $\Lambda\alpha$ potentials represented by a Gaussian form
\begin{equation}
\label{eq:Gauss}
    U_{\Lambda \alpha} =
    \sum_i U_i \exp\left[{-\left(\dfrac{r}{a_i}\right)^2}\right],
\end{equation}
where $U_i$ and $a_i$ are the potential parameters.
Chi3 is the Skyrme-type potential~\cite{Jinno:2023xjr} constructed by reproducing the results of the $\Lambda$ single-particle potential with the $YN$ and $YNN$ interactions from chiral effective field theory (chiral EFT)~\cite{Gerstung:2020ktv}, while LY-IV is the conventional Skyrme-type potential~\cite{Lanskoy:1997xq}.
It is important to distinguishing these potentials toward solving the hyperon puzzle of neutron stars:
Chi3 is strongly repulsive at high densities and $\Lambda$'s are suppressed in neutron star, while LY-IV is attractive, resulting in the softening of the equation of state by the appearance of $\Lambda$'s in neutron stars.
The function forms of the single-particle potential $U_\Lambda$ and the effective mass $m_\Lambda^*$ in the Skyrme-type potential are given as
\begin{align}
    U_\Lambda(\bm{r}) &=  a^\Lambda_1\rho_N(\bm{r}) + a^\Lambda_2 \tau_N(\bm{r}) - a^\Lambda_3 \Delta \rho_N(\bm{r}) 
    + a^\Lambda_4 \rho^{4/3}_N(\bm{r}) + a^\Lambda_5 \rho^{5/3}_N(\bm{r}), \label{eq:ULambda} \\
    \dfrac{1}{2m_\Lambda^*(\bm{r})} &= \dfrac{1}{2 m_\Lambda} + a^\Lambda_2 \rho_N(\bm{r}), 
    \label{eq:mLambdastar}    
\end{align}
where $a^\Lambda_i$ is the potential parameter and $m_\Lambda$ is the mass of $\Lambda$ in vacuum.
The nucleon density $\rho_N$ and kinetic density $\tau_N$ measured from the center-of-mass~\cite{Jinno:2023xjr} are calculated by using the wave function of $\alpha$ expressed by the products of nucleon Gaussian as~\cite{Akaishi:1986gm}
\begin{align}
     \label{eq:AlphaWF}
    \psi_\alpha(\bm{r}_1,\cdots,\bm{r}_4)&=\prod^{4}_{i=1} \phi(\bm{r}_i), \\
    \phi(\bm{r}) &= \left(\dfrac{2\nu}{\pi}\right)^{3/4} \exp[-\nu \bm{r}^2],
\end{align}
with the width parameter $\nu=0.20~\text{fm}^{-2}$ is chosen to reproduce the experimental value of the $\alpha$ charge radius $\sqrt{\langle \bm{r}^2 \rangle_A} = 1.68~\text{fm}$~\cite{Krauth:2021foz}.
Parameters $a^\Lambda_i$ for $i=1,2,4,$ and $5$ are same as those used for the hypernuclear calculation~\cite{Jinno:2023xjr}, while $a^\Lambda_3$ is refitted to reproduce the experimental binding energy of ${}^5_\Lambda {\rm He}$.
It has been shown that both Chi3 and LY-IV models reproduce the experimental data of medium to heavy hypernuclei~\cite{Jinno:2023xjr} by tuning $a^\Lambda_3$, which does not change the potential in uniform matter.

\begin{figure}[t]
\centering
\includegraphics[scale=0.5]{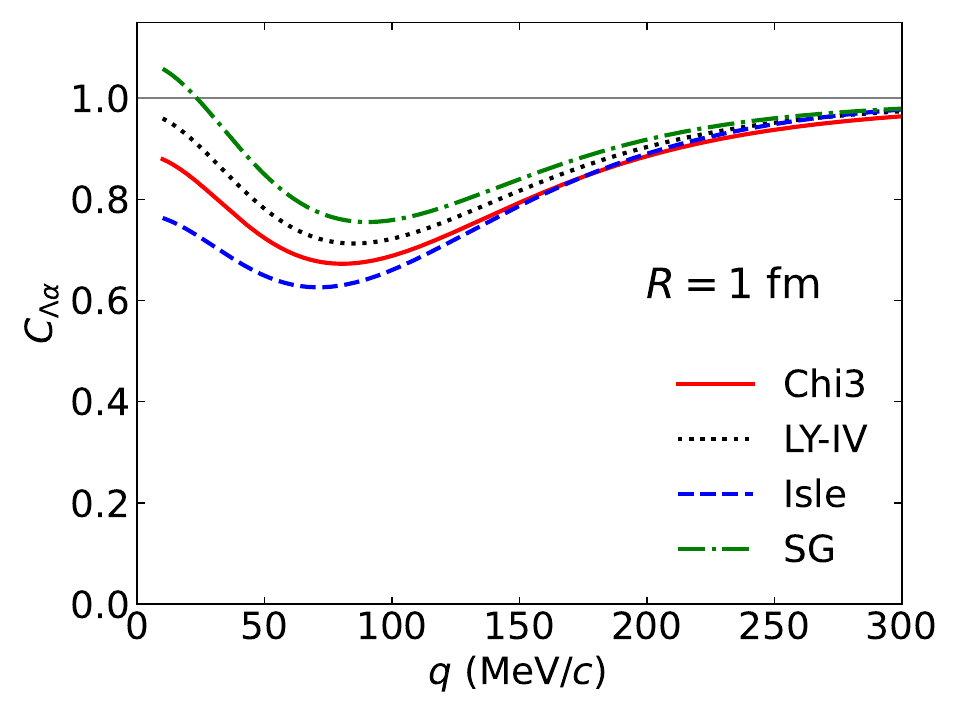}
\caption{$\Lambda\alpha$ correlation functions for a source size $R=1~\text{fm}$.
The results with Chi3, LY-IV, Isle, and SG potentials are represented by solid, dotted, dashed, and dash-dotted lines, respectively.
Figure is adopted from \cite{Jinno:2024tjh}.}
\label{fig:LamAlphaCF}
\end{figure}

The $\Lambda \alpha$ potentials are shown in the left panel of Fig.~\ref{fig:LamAlphaPot}.
They have different behavior at short range, reflecting the properties of $\Lambda$ at high densities.
In contrast, they have similar tail at the longer range,
where the $\Lambda N$ interaction is well described by the two-pion exchange at long distance.
In all cases, the interaction ranges are of the order of $2$-$3~\text{fm}$.
In the right panel of Fig.~\ref{fig:LamAlphaPot}, the reduced effective mass $\mu^* = m^*_\Lambda m_\alpha / (m^*_\Lambda + m_\alpha)$, which appears in the $\Lambda \alpha$ Schr\"odinger equation, is shown. Although the effective mass reduces of about $300~\text{MeV}$ at $r=0$ in Chi3, we have checked that the difference between the models with and without the effective mass does not affect the $\Lambda\alpha$ correlation function if $a^\Lambda_3$ is readjusted~\cite{Jinno:2024tjh}.

In Fig.~\ref{fig:LamAlphaCF}, we show $\Lambda\alpha$ correlation functions calculated by the KP formula~\eqref{eq:KPformula} for a source size $R=1~\text{fm}$.
The differences among the models are manifest in the lower momentum region.
The correlation functions are suppressed in the order of the strength of the $\Lambda\alpha$ potential at short range.
This type of suppression is typical for standard repulsive potentials: the repulsive potential reduces the absolute value of the wave function at short range, leading to the suppression in the correlation function.
These results suggest that precise measurements of the $\Lambda\alpha$ correlation function can distinguish the short range behavior of the $\Lambda\alpha$ potential, which is a key to discuss the internal structure of neutron stars.

\section{$\Xi \alpha$ momentum correlation}
\label{sec:XiAlpha}

The $\Xi N$ potentials have four components in the $s$-wave channel, ${}^{11}S_0$, ${}^{31}S_0$, ${}^{13}S_1$, and ${}^{33}S_1$ where the notation ${}^{2I+1,2s+1}L_J$ is adopted with the spin $s$, isospin $I$, and the total angular momentum $J$.
From the analysis of the $p\Xi^{-}$ correlation function~\cite{ALICE:2019hdt,ALICE:2020mfd} based on the HAL QCD potentials~\cite{Kamiya:2019uiw} and chiral EFT~\cite{Haidenbauer:2018jvl,Haidenbauer:2022esw,Haidenbauer:2022suy}, the strength of the ${}^{11}S_0$ attraction is not strong enough to make a $\Xi N$ bound state.
Studying the $\Xi \alpha$ potential can reveal further nature of the $\Xi \alpha$ potentials.
The $\Xi\alpha$ interaction is expressed by the weighted sum of the $\Xi N$ interaction components as 
$ \left[ V({}^{11}S_0)+3V({}^{31}S_0)+3V({}^{13}S_1)+9V({}^{33}S_1)\right]/16$.
Namely, the $\Xi\alpha$ interaction is dominated by the ${}^{33}S_1$ component of the $\Xi N$ interaction~\cite{Hiyama:2022jqh}.
Thus the $\Xi\alpha$ correlation brings the information complementary to the $p\Xi^-$ correlation.

\begin{figure}[t]
	\centering
        \includegraphics[scale=0.6]{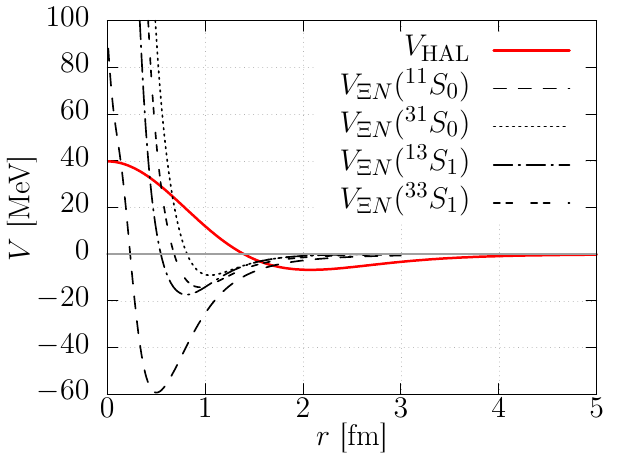}
	\includegraphics[scale=0.6]{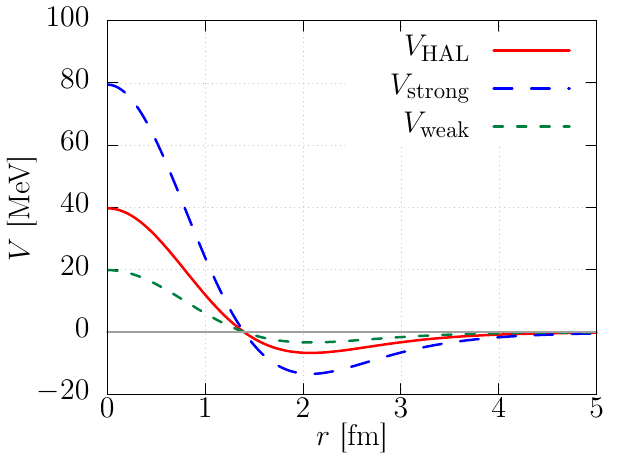}
	\caption{(left panel) Folding $\Xi\alpha$ potential~\cite{Hiyama:2022jqh} and HAL QCD $\Xi N$ potentials~\cite{HALQCD:2019wsz}.
    (right panel) Folding $\Xi \alpha$ potential based on HAL QCD $\Xi N$ potentials and its variations. Narrow and wide dashed lines represent the strong and weak models whose potentials are twice and half of $V_{\rm HAL}$, respectively.
    Figures are adopted from Ref.~\cite{Kamiya:2024diw}.}
	\label{fig:XiAlphaPot}
\end{figure}

We employ the folding $\Xi \alpha$ potential~\cite{Hiyama:2022jqh} based on the HAL QCD $\Xi N$ potentials~\cite{HALQCD:2019wsz}.
The $\Xi N$-$\Lambda\Lambda$ coupling is effectively renormalized into the $\Xi\alpha$ single-channel potential, and then the potential has no imaginary part.
The folding potential is parametrized by the sum of the 20 Gaussians as
\begin{align}
	V_{\rm HAL}(r)= \sum_{i=1}^{20}V_i \exp(-\nu_i r^2), \label{eq:V_Hiyama}
\end{align}
where $V_i$ and $\nu_i$ are the parameters.
The $\Xi \alpha$ folding potentials and the HAL QCD $\Xi N$ potential are shown in the left panel of Fig.~\ref{fig:XiAlphaPot}. The potential $V_{\rm HAL}$ has repulsive core at the short range and the broad pocket of the waek attraction within $r \approx 1.5$-$4.0~\text{fm}$. The potential has larger interaction range of $4~\text{fm}$ than the $\Lambda\alpha$ potential in Fig.~\ref{fig:LamAlphaPot} owing to the one-pion exchange contribution, which is prohibited for the $\Lambda \alpha$ system.

For the $\Xi^{-}\alpha$ system, the further attraction by the Coulomb interaction acts as
\begin{align}
	V_{\rm Coulomb}(r) = -\dfrac{2\alpha}{r},
    \label{eq:V_Coulomb}
\end{align}
with the fine structure constant $\alpha$.
By solving the Schr\"odinger equation with $V_{\rm HAL}+V_{\rm Coulomb}$, we find a shallow bound state $^{5}_{\Xi} {\rm H}$ with the binding energy $B=0.47~\text{MeV}$.

From chiral EFT~\cite{Le:2021gxa}, the $^{5}_{\Xi} {\rm H}$ bound state with $B=2.16~{\rm MeV}$ is found by employing next-to-leading order (NLO) $\Xi N$ interactions.
This indicates that the chiral $\Xi\alpha$ interaction is more attractive than the HAL QCD based potential $V_{\rm HAL}$.
On the other hand, there remains the possibility that the $\Xi\alpha$ potential might be less attractive than $V_{\rm HAL}$ where the bound state is mainly generated by the Coulomb interaction. To examine the theoretical uncertainty in the $\Xi\alpha$ potential, we consider two variations of the potential; $V_{\rm strong}=2V_{\rm HAL}$ and $V_{\rm weak}=V_{\rm HAL}/2$ for the stronger and weaker potentials, respectively.
These variations are shown in the right panel of Fig.~\ref{fig:XiAlphaPot}.
With $V_{\rm strong}+V_{\rm Coulomb}$, the binding energy of $\Xi^- + \alpha$ system is obtained as $2.08~\text{MeV}$, the value obtained from chiral EFT~\cite{Le:2021gxa}. With $V_{\rm weak}+V_{\rm Coulomb}$, a shallow bound state with $B=0.18~{\rm MeV}$ is found.
For $\Xi^0 \alpha$ system, only $V_{\rm strong}$ exhibits the bound state.
The binding energies of the $\Xi^- \alpha$ and $\Xi^0 \alpha$ systems are summarized in Table~\ref{tab:bd_alphaXi}.

\begin{table}
        \caption{Binding energies of the $\Xi^-\alpha$ and $\Xi^0\alpha$ systems. The Coulomb interaction~\eqref{eq:V_Coulomb} is included in the $\Xi^-\alpha$ calculation. }
		\label{tab:bd_alphaXi}
%\begin{ruledtabular}
        \centering
		\begin{tabular}{ccc}
            \hline
			Potential&$\Xi^-\alpha$ [MeV] & $\Xi^0 \alpha$ [MeV]\\
			\hline
			$V_{\rm HAL}$ & $0.47$  & - \\
			$V_{\mathrm{strong}} $ & $2.08$   &  1.15 \\
			$V_{\mathrm{weak}} $ & $0.18$  & -- \\
            \hline
		\end{tabular}
%\end{ruledtabular}
\end{table}

\begin{figure}[t]
	\centering
	\includegraphics[scale=0.6]{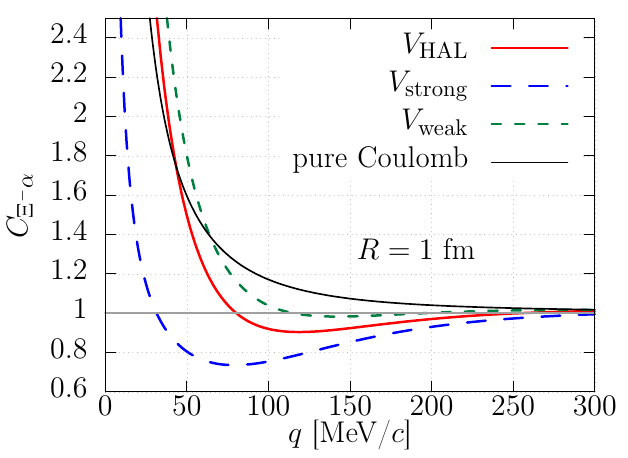}
	\includegraphics[scale=0.6]{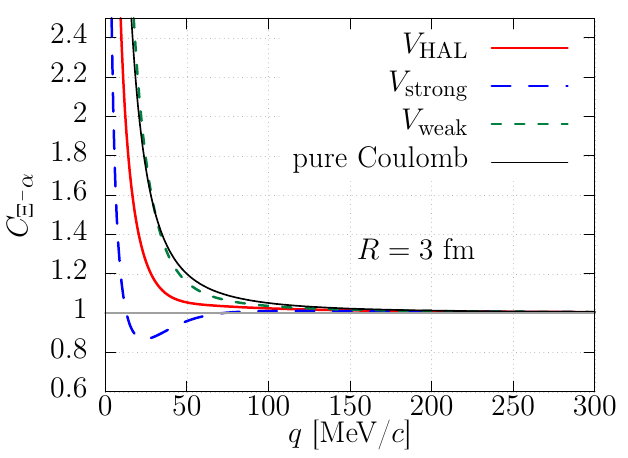}
	\caption{$\Xi^- \alpha$ correlation as functions of the relative momentum $q$ for the source sizes of $1$ and $3~\text{fm}$. The thick solid line corresponds to the model using the folding potential employing the HAL QCD $\Xi N$ potentials~\cite{HALQCD:2019wsz}.
    The narrow and wide dashed lines represent the strong and weak models, which are half and twice $V_{\rm HAL}$, respectively.
    The thin solid line corresponds to the model with only Coulomb interaction.
    Figures are adopted from Ref.~\cite{Kamiya:2024diw}.}
	\label{fig:XiAlphaCF}
\end{figure}

In Fig.~\ref{fig:XiAlphaCF}, the correlation functions calculated by the KP formula~\eqref{eq:KPformula} are shown for the source size $R=1$ and $3~\text{fm}$.
For the source size of $1~\text{fm}$, the folding potential~\eqref{eq:V_Hiyama} plus Coulomb potential $V_{\rm HAL}+V_{\rm Coulomb}$ exhibits bump structure around the relative momentum $q \approx 50$-$200~\text{MeV}/c$ compared to the pure Coulomb case.
The weak potential model has same tendency as the $V_{\rm HAL}+V_{\rm Coulomb}$ model but with smaller suppression.
These bump structure is a typical behavior when the potential exhibits a deeper bound state compared to the pure Coulomb case.
The $V_{\rm strong}+V_{\rm Coulomb}$ model has the largest bump structure,
reflecting the large repulsive core in the relative distance $r<1~\text{fm}$ as discussed in Ref.~\cite{Kamiya:2024diw}.
These tendencies are also seen in the results of the larger source size of $3~\text{fm}$:
the $V_{\rm weak}+V_{\rm Coulomb}$ model is almost same as the pure Coulomb model,
the $V_{\rm HAL}+V_{\rm Coulomb}$ model exhibits moderate suppression,
and the $V_{\rm strong}+V_{\rm Coulomb}$ model has strong suppression in the low momentum region.
Therefore, the source size dependence on the correlation function reflects the difference in the bound state property and the repulsive core of the potentials.

\section{Summary}
\label{sec:summary}

We have investigated the $\Lambda\alpha$ and $\Xi\alpha$ momentum correlations, which can be measured in the future experiments of high-energy collisions.
First, we have examined the $\Lambda\alpha$ system by comparing four models:
Two of them are the conventional Gaussian potentials.
The others are newly constructed from the Skyrme-type $\Lambda$ single-particle potentials with different behaviors in the high density region.
For Skyrme-type potentials, one coefficient for the term involving the derivative of the nucleon density is tuned to reproduce the experimental data of the binding energy of ${}^5_{\Lambda}{\rm He}$.
The difference among the potentials lies in their property at short range, ranging from attractive to repulsive, while the binding energy of ${}^5_\Lambda {\rm He}$ and the long-range behavior are similar.
The calculated correlation functions show differences for a small-size source, implying that the unresolved short-range part of the $\Lambda N$ interaction and the $\Lambda$ single-particle potential in dense nuclear matter may be revealed by future precise measurements of the $pp$ or $p$-nucleus collisions.

Second, we have investigated the $\Xi\alpha$ system by employing the folding $\Xi\alpha$ potential~\cite{Hiyama:2022jqh} based on the HAL QCD $\Xi N$ potentials~\cite{HALQCD:2019wsz}.
To examine the sensitivity of the momentum correlation to the bound state nature of the $\Xi \alpha$ system, we prepared two variations, referred to as the strong and weak potentials, by multiplying the folding potential by $2$ and $1/2$, respectively.
The strong potential model has a $\Xi^- \alpha$ bound state similar to the result from chiral EFT~\cite{Le:2021gxa}.
These three models exhibit different source size dependence in the $\Xi^-\alpha$ correlation function,
reflecting the bound state properties and the short range behavior of the $\Xi\alpha$ potential.
Notably, difference among models is found also for a source size of $3~\text{fm}$, where the calculated $\Lambda \alpha$ correlation functions are almost identical.

In this study, we consider the small source sizes of $R=1~\text{fm}$,
which is the typical size used in the analysis of the two-hadron correlation functions obtained in the high-energy $pp$ collisions, assuming that $\alpha$ is created at the initial fireball.
However, the effective source size of the composite particle $\alpha$, which can also be formed by the coalescence of four nucleons emitted from the fireball, must be larger than those with single hadron emissions~\cite{Mrowczynski:2019yrr,Bazak:2020wjn,Mrowczynski:2021bzy}.
In fact, the $\alpha$ particle is treated as a point-like particle in the Koonin-Pratt formula~\eqref{eq:KPformula}.
For a more realistic study of the $Y \alpha$ correlation, the five-body scattering problem of four nucleons and alpha should be solved to take into account the coalescence effect.

Measurements of $Y\alpha$ correlation functions may be realized by the collisions with the center-of-mass energy per nucleon pair of $\sqrt{s_{NN}}<10~\text{GeV}$, where many $\alpha$ particles would be produced according to Ref.~\cite{Andronic:2010qu}.
Then, the $Y\alpha$ correlation function can be measured in the future facilities such as FAIR, NICA, HIAF, and J-PARC HI to elucidate the $Y\alpha$ potential and reveal further nature of the $YN$ interactions.

\bibliographystyle{elsarticle-num}
\bibliography{ref}

\end{document}